%% file: main.tex
\documentclass[sigconf,authorversion]{acmart}

\usepackage{subcaption}
\usepackage{multirow}
\usepackage{algorithm}
\usepackage{algorithmic}
\usepackage{amsmath}

\usepackage{amssymb}
\usepackage{rotating}
\usepackage{makecell}

\newcommand{\NA}{---}

\AtBeginDocument{%
  }

\copyrightyear{2025}
\acmYear{2025}
\acmDOI{XXXXXXX.XXXXXXX}
\acmISBN{978-1-4503-XXXX-X/2026/05} 
\acmConference[SenSys '26]{International Conference on Embedded Artificial Intelligence and Sensing Systems}{May 11--14, 2026}{Saint-Malo, France}

\acmSubmissionID{223}

\begin{document}

\title[HERMES: Heterogeneous Edge Realtime Measurement and Execution System]{HERMES: A Unified Open-Source Framework for Realtime Multimodal Physiological Sensing, Edge AI, and Intervention in Closed-Loop Smart Healthcare Applications}

\author{Maxim Yudayev}
\orcid{0000-0001-9521-1537}
\affiliation{%
  \institution{KU Leuven}
  \city{Leuven}
  \country{Belgium}
}
\email{maxim.yudayev@kuleuven.be}

\author{Juha Carlon}
\orcid{0009-0003-3178-1153}
\affiliation{%
  \institution{KU Leuven}
  \city{Leuven}
  \country{Belgium}
}
\email{juha.carlon@kuleuven.be}

\author{Diwas Lamsal}
\orcid{0009-0006-1125-2299}
\affiliation{%
  \institution{KU Leuven}
  \city{Leuven}
  \country{Belgium}
}
\email{diwas.lamsal@kuleuven.be}

\author{Vayalet Stefanova}
\orcid{0009-0006-0137-5033}
\affiliation{%
  \institution{KU Leuven}
  \city{Leuven}
  \country{Belgium}
}
\email{vayalet.stefanova@kuleuven.be}


\author{Benjamin Filtjens}
\orcid{0000-0003-2609-6883}
\affiliation{%
  \institution{Delft University of Technology}
  \city{Delft}
  \country{The Netherlands}
}
\email{b.filtjens@tudelft.nl}

\renewcommand{\shortauthors}{M. Yudayev et al.}

\begin{abstract}
Intelligent assistive technologies are increasingly recognized as critical daily-use enablers for people with disabilities and age-related functional decline.
Longitudinal studies, curation of quality datasets, live monitoring in activities of daily living, and intelligent intervention devices, share the largely unsolved need in reliable high-throughput multimodal sensing and processing.
Streaming large heterogeneous data from distributed sensors, historically closed-source environments, and limited prior works on realtime closed-loop AI methodologies, inhibit such applications.
To accelerate the emergence of clinical deployments, we deliver HERMES - an open-source high-performance Python framework for continuous multimodal sensing and AI processing at the edge.
It enables synchronized data collection, and realtime streaming inference with user PyTorch models, on commodity computing devices.
HERMES is applicable to fixed-lab and free-living environments, of distributed commercial and custom sensors.
It is the first work to offer a holistic methodology that bridges cross-disciplinary gaps in real-world implementation strategies, and guides downstream AI model development.
Its application on the closed-loop intelligent prosthesis use case illustrates the process of suitable AI model development from the generated constraints and trade-offs.
Validation on the use case, with 4 synchronized hosts cooperatively capturing 18 wearable and off-body modalities, demonstrates performance and relevance of HERMES to the trajectory of the intelligent healthcare domain.
\end{abstract}



\input{figures/teaser}

\received{dd-MMM-yyyy}
\received[revised]{dd-MMM-yyyy}
\received[accepted]{dd-MMM-yyyy}

\maketitle



\section{Introduction}
\label{sec:introduction}
\input{content/01-intro}

\section{Related Works}
\label{sec:related}
\input{content/02-related}

\section{HERMES}
\label{sec:hermes}
\input{content/03-methodology}

\input{content/04-architecture}

\input{content/05-evaluation}

\section{Discussion}
\label{sec:discussion}
\input{content/06-discussion}

\section{Limitations and Future Work}
\label{sec:limitations_and_future}
\input{content/07-limitations}

\section{Conclusion}
\label{sec:conclusion}
\input{content/08-conclusion}

\begin{acks}
This study was funded, in part, by the AidWear project funded by the Federal Public Service for Policy and Support, 
the AID-FOG project by the Michael J. Fox Foundation for Parkinson’s Research under Grant No.: MJFF-024628,
the strategic basic research project RevalExo (S001024N) funded by the Research Foundation Flanders,
and the Flemish Government under the Flanders AI Research Program (FAIR).
\end{acks}

\bibliographystyle{ACM-Reference-Format}
\bibliography{bibliography}
\end{document}

%% file: figures/teaser.tex
\begin{teaserfigure}
  \centering
  \includegraphics[width=0.85\textwidth]{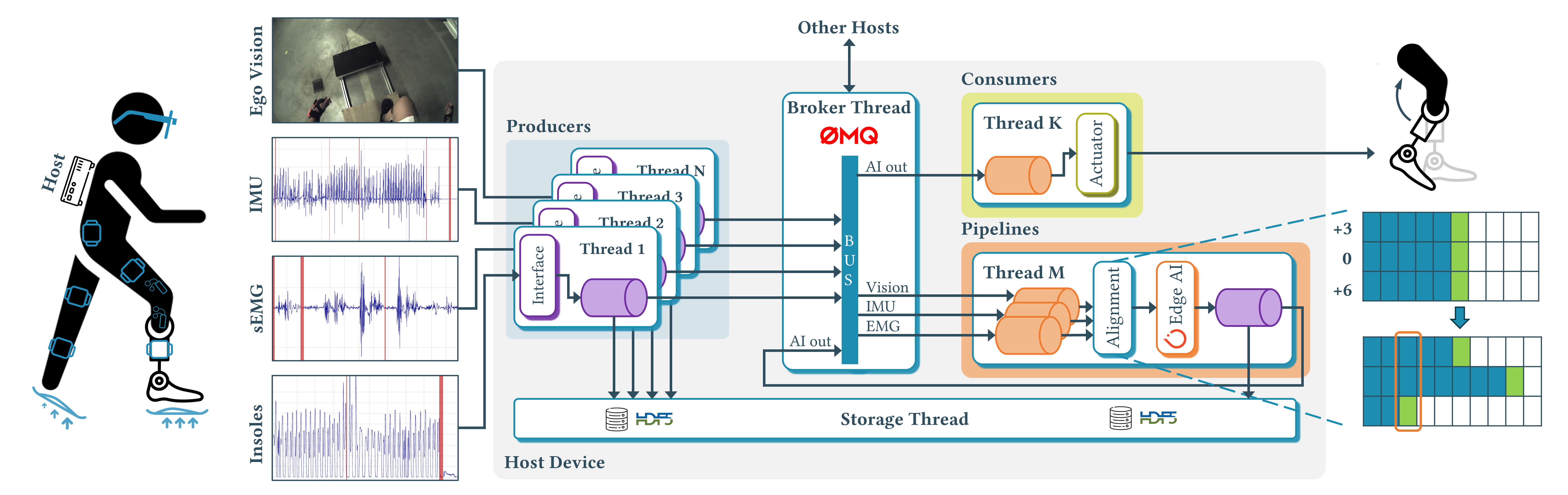}
  \caption{Event-driven system architecture on an intelligent prosthesis usecase, detailing the asynchronous flow of data, subject to missingness (red), between different system components producing or consuming data, or both, across distributed hosts.}
  \Description{A pipeline-like modular multi-threaded processing architecture on one of many similarly functioning interconnected distributed host devices.
  The visualized system is hosted on the wearable minicomputer that an amputee is wearing on their back.
  Heterogeneous wearable physiological sensors are connected to the host and capture streaming data in realtime, each having occasional missing data due to external factors or sensor's internal errors.
  In total, four different types of sensors are connected to the system: smartglasses for eye tracking and egocentric vision, inertial sensors, surface electromyography sensors, and foot pressure sensors.
  The captured samples of a sensor trigger event-based processing by the corresponding producer thread inside the HERMES system.
  Once placed by each thread in a specialized datastructure, the data is dynamically passed to the broker thread for communicating it to any local or remote subscribers interested in the data of the corresponding sensor.
  All data entering the system is safely stored in the efficient HDF5 file format by the host using a storage thread.
  The host has one consumer thread and one pipeline thread, that do some local processing on the captured data in realtime.
  The pipeline thread is subscribed to vision, inertial and electromyography data for AI-based realtime movement intent prediction using a custom PyTorch model.
  Everytime, new data is available, it is passed to the PyTorch AI model through an alignment block that uses a selected alignment policy to preprocess samples for the model's input.
  In the current scenario, the block aligns samples across sensors to get rid of different propagation delays for samples to correctly correlate to the same instance in time.
  The computed predictions are then published back to the broker thread that routes the predictions to the consumer thread, which uses that processed data to wirelessly control the actuation of the prosthesis to bend the knee.}
  \label{fig:teaser}
\end{teaserfigure}

%% file: content/01-intro.tex
Practical realtime multimodal edge AI innovations require far more than a lightweight model running on a constrained device.
Bringing AI to the edge in the real-world demands continuously acquiring and aggregating a variety of data sources - modalities, from a distributed and heterogeneous set of interconnected sensors and host devices.
These modalities operate independently from one another in a continuously streaming fashion.
Aggregating them reliably is a challenge: information is produced asynchronously, across various wired and wireless connections, over distributed hosts, at different sampling frequencies, number of dimensions, and levels of reliability \cite{garg-vision-acm-2010, ortiz-dataforai-jmir-2024}.

Previous works on edge AI systems addressed several of these aspects at a time, which we group into data synchronization \cite{wolling-timediscrepancy-iswc-2025, trollmann-timediscrepancy-thesis-2025, wu-multimodalsensingdemo-sensys-2025, ni-multimodalwearablesurvey-arxiv-2024, kolakowska-dtwsynchronization-sensys-2025, ma-nymeria-eccv-2024}, data reliability \cite{ferrara-wearablellm-sensors-2024}, data fusion \cite{acosta-multimodalai-nature-2022, li-multimodalsurvey-arxiv-2024, tsepa-multimodalprosthesis-icra-2023, wang-multimodalfusion-informationfusion-2025}, AI model optimization \cite{tsepa-multimodalprosthesis-icra-2023, just-mlprosthesis-ieeeaccess-2024, yudayev-rtstgcn-mlsp-2024}, acquisition-to-deployment performance degradation \cite{vela-temporaldegradation-nature-2022, ding-biosignalllm-arxiv-2024, narayanswamy-foundationwearablemodels-arxiv-2024}, real-world deployment \cite{wu-asyncmultimodalinference-sensys-2024, delpreto-actionsense-neurips-2022, strandquist-athomecollection-jove-2023}, hyper-specialized embedded device development \cite{wiese-multimodaliot-arxiv-2025, mahato-hybridsensors-nature-2024, ates-sensortrends-nature-2022},
foundation models \cite{ansari-chronos2-arxiv-2025, langer-opentslm-arxiv-2025, xu-lsm-arxiv-2025, ding-biosignalllm-arxiv-2024, narayanswamy-foundationwearablemodels-arxiv-2024}.
Undertaking each in isolation has implications on orthogonal design challenges, which hinders practical solutions \cite{diraco-review-sensors-2023, heydari-tinyml-sensors-2025, ping-multimodalfusion-jiot-2025}.
Lack of cohesive methodologies for addressing heterogeneity of multimodal data, and of real-world centered operation strategies, reveals a largely understudied area that throttles efforts in intelligent health innovations \cite{gu-biosignalsfoundationsurvey-techrxiv-2025}.

Developing such real-world systems is a challenge at the intersection of a breadth of technical domains.
It requires expertise in sensors and electronics interfacing, computer networks, device synchronization, distributed realtime software, IT administration, hardware-aware AI model design, system performance evaluation, data aggregation and alignment, data analysis, system-level integration.
Many crucial technical challenges outside of one's research scope are involved in enabling testing of a novel edge AI technique, or in validating a hypothesis in a closed-loop clinical use case.

Selection and tuning of an AI algorithm for deployment is frequently done via an iterative process or a neural architecture search \cite{gao-nas-taco-2025, akhauri-naslatency-mlsys-2024}.
A model is deployed in the wild or an emulated setting \cite{gill-edgereview-springer-2025} to validate its realtime processing capability, suitability for the application constraints, and sensitivity to the real-world nature of sensed signals.
Lack of prior, real-world informed, design guidelines - like permitted inference time budget, stochastic nature of the signals, inhibits cohesive implementation for AI researchers.
Researchers rely on intuition and previous experiences to choose models, and later attempt to compress them to fit the apparent constraints of production edge systems \cite{hohman-aidesign-chi-2024, bringmann-edgeaicodesign-codes-2021}.
The results require iterative model adaptations or ablation studies to strike balance between constraints and performance.

Real-world factors like sensor data transmission delays, synchronization skew and drift, system's internal processing delays, sensor packet loss rate - are all not known before a system is implemented.
These factors impact data reliability, synchronized sensing of distributed modalities, and throughput of the implementation.
Inadequate execution causes cascading effects on downstream AI models and falls outside the main scope of one's exploration.
Bypassing the challenges requires a flexible, validated, and efficient high-performance implementation, and unified methodologies for holistic system design. 

We demonstrate that treating complex real-world edge AI deployments as a holistically designed system offers a reductionist view on the challenges.
The view disentangles connected design challenges and contextualizes overall system architecture with respect to the deployment objective.
The resulting architecture equips researchers with a flexible, efficient, high-performance implementation, which delivers deep insights into the practical real-world aspects of an end-to-end system.
These insights offer clear lasting guidelines and stimulate the development of realtime AI models that meet the constraints of the target application.

In this work, we formulate a methodology in the specification of such systems, following a top-down approach from the set of our generalized multimodal realtime continuous inference strategies.
The formulation catalyzes definition of a set of clear guidelines and constraints, which aid the development of practical target systems.
We show how following this methodology in the design of an end-to-end system fulfills real-world requirements of healthcare applications and streamlines closed-loop AI model development.
We complete our work with the introduction of \verb|HERMES|\footnote{\anon[\url{https://anonymous.4open.science/r/sensys26_223}]{\url{https://github.com/maximyudayev/hermes}}}, an open-source developer-friendly high-throughput Python framework.
It embodies the proposed methodology to bridge the domain gaps, empowers users with a unified reliable platform that de-risks system integration challenges, and allows to focus on the core research and development objectives.

The novel contributions of this work are summarized as:
\begin{itemize}
\item A methodology for contextualizing a continuous multimodal realtime edge AI use case.
\item A real-world focused technique for continuous synchronization of distributed modalities in a streaming system.
\item A validated open-source high-performance Python framework with PyTorch integration for physiological sensing and AI processing of smart healthcare applications.
\end{itemize}

%% file: content/02-related.tex
\input{tables/system_requirements}

A range of emerging intelligent health- and med-tech applications rely on or require continuous streaming and processing of the ubiquitous multimodal data to drive personalized realtime closed-loop interventions, analytics, and performance monitoring \cite{ates-sensortrends-nature-2022}.
This data ranges from physiological wearable sensors, to telemetry data from actuated assisted devices, to egocentric and exocentric video \cite{ma-nymeria-eccv-2024, chadwell-prosthesismonitor-jner-2020, mancini-pdtechnologies-jpd-2025}.
The sensors are spread across various stationary and wearable acquisition devices connected wired and wirelessly, in a computer network.
With the changing human-centered emphasis in healthcare on preventative therapy and rehabilitation \cite{claeys-exoskeletonperspectives-wearabletech-2025, kooij-exoreview-science-2025}, and on improvement of the quality of life with a chronic condition \cite{mancini-pdtechnologies-jpd-2025}, research in always-on and long-running real-world systems outside of fixed lab environments started receiving more attention and importance \cite{chadwell-prosthesismonitor-jner-2020}.
However, technical multi-disciplinary challenges of continuous multimodal sensing and processing \cite{parajuli-realtimeprosthesisreview-sensors-2019} and the lack of extendable open-source research frameworks that address them \cite{kooij-exoreview-science-2025} cause large solutions disparity \cite{heydari-tinyml-sensors-2025} and constrict development, evaluation, and valorization of clinically impactful innovations.

The challenges hinder applicability of novel methods to applications in free-living, realtime continuous intelligence, and reliable high-quality data collection, inhibiting advances in personalized wearable technologies across use cases \cite{yang-imufog-jner-2024, chadwell-prosthesismonitor-jner-2020, tsepa-multimodalprosthesis-icra-2023, ferrara-wearablellm-sensors-2024, ortiz-dataforai-jmir-2024}.
Management and collection of heterogeneous modalities, consistent and aligned high-quality sensor data, realtime processing, interoperability of systems and methods in diverse environments, integration of sensing and AI processing, and lack of infrastructure for free-living sensing and data collection remain major unresolved challenges \cite{li-multimodalsurvey-arxiv-2024, ferrara-wearablellm-sensors-2024, ping-multimodalfusion-jiot-2025, parajuli-realtimeprosthesisreview-sensors-2019, tsepa-multimodalprosthesis-icra-2023, kooij-exoreview-science-2025}.
Despite significant positive influence of multiple physiological modalities on AI prediction performance \cite{yang-multimodalfog-tnsre-2025}, no clear guidelines exists to delineate the model development workflow \cite{hohman-aidesign-chi-2024}.

To address the multi-disciplinary challenges that the field faces, a holistic approach to the system architecture is required \cite{diraco-review-sensors-2023}.
The approach should be designed with the nature of real physiological signals in mind \cite{xu-lsm-arxiv-2025} - missingness, asynchronism, heterogeneity, delay between sensing and manifestation of a signal.
Several notable works have been developed that overlap the scope of the present study.
Table~\ref{tab:challenges} lists the most relevant and established of them against the crucial requirements, to underline the gaps in the state-of-the-art systems.

LabView \cite{elliott-labview-jala-2007} is a standard industrial sensing and visual programming platform with a diverse hardware ecosystem.
It comes at the cost of a high learning curve, expensive proprietary hardware and software licenses, and vendor lock-in.
Vicon \cite{vicon} visual motion capture system has been the cornerstone of synchronous multi-sensor physiological data acquisition for human movement analysis and visual effects since 1984.
It is a reliable system that abstracts technical challenges of synchronous physiological sensing away from the user via a simple non-technical interface.
Its main downsides are constraint to the fixed lab environment, limited support to a subset of commercial sensors and prebuilt analysis toolkits, lack of custom closed-loop AI processing, and requirement for the dedicated proprietary acquisition server.
Xsens MVN Analyze \cite{xsens} is a ubiquitous IMU-based motion capture system in the entertainment and movement sciences domain.
It offers data recording and streaming to external network systems for integration with 3rd parties.
The system is a proprietary software-hardware bundle with a high cost, limited integration with other sensors, and in our experience is not suitable for long continuous recordings.

In distributed web and IoT applications, message queue (MQ) systems, Tab.~\ref{tab:challenges} (row 6-7), have been a popular middleware choice for distributed message streaming and processing applications \cite{fu-mqsystems-ieee-2021}.
NATS \cite{nats} stands out thanks to its diverse out-of-the-box functionality and built-in support for the common distributed communication patterns: load balancing, message brokering, dynamic service discovery, canary deployment.
Middleware like RabbitMQ, Kafka, NATS are fault-tolerant high-performance microservices targeted at distributed web applications.
MQ systems are a centralized microservice on dedicated server, relaying data between connected clients and other microservices to construct complex applications.
They were designed and tuned for web applications and are too high in the abstraction level for high-performance low-latency realtime edge use cases, where lower-level middleware for inter- and intra-process communication on the local network is required \cite{ubicoders-middleware-youtube-2025}.
MQTT is a popular lightweight IoT protocol for distributed exchange of small data.
It offers a three level quality-of-service (QoS) message exchange through a centralized message broker. 
Because of it's simplicity, it is compatible with simple embedded electronics, but is unsuitable for high-throughput systems.

In 2012, LSL \cite{kothe-lsl-biorxiv-2024} was built to address the needs of the neurophysiological research community in a robust open-source framework for accurate synchronized multi-channel neuro-signal acquisition.
The development enabled alignment of separate neural streams to the millisecond level and empowered a breadth of research \cite{wang-lslscope-springer-2023}.
Because LSL was designed for synchronous recording and for neural signals, it does not provide a stream processing capability for custom AI models, support for video data Tab.~\ref{tab:challenges}(3), or out-of-the-box visualization Tab.~\ref{tab:challenges}(4).
LSL aligns streams inside its core middleware by compensating for the difference in propagation delays from each sensor, based on the estimate thereof and on a manual offset parameter.
The fixed internal alignment strategy adds latency and limits the flexibility to alternative alignment methods.

Several novel works built on top of LSL to address its limitations.
CLAID \cite{langer-claid-fgcs-2024} built a live smartphone-based data collection and processing system that offloads ML computation, Tab.~\ref{tab:challenges}(12), to the cloud or an edge server for analysis of the wearable multimodal data.
Most recently, \cite{elmakrini-physiosense-ram-2025} brought multimodal data collection for workspace human movement research outside the lab as a mobile sensing backpack, and used sensors that were previously limited to the fixed lab environment.
LSL-based systems face the same shortcomings, Tab.~\ref{tab:challenges}(6), when moving to realtime streaming - inflexibility in fusion strategy and in triggering of computation on partial data availability, both of which an expected prerequisite for downstream AI applications \cite{wu-asyncmultimodalinference-sensys-2024}.

In the domain of in-vivo neurophysiology, Tab.~\ref{tab:challenges}(2), Syntalos \cite{klump-syntalos-nature-2025} proposed an extensive software system for precisely aligned heterogeneous sensing and closed-loop interventions for mice experiments.
Continuous synchronization between sensors ensured clock skew did not compound, but large jitter in synchronization of up to 40ms was observed.
Dedicated parallel sensing systems in low-level programming languages \cite{newman-multimodalneuralsystem-nature-2025, klump-syntalos-nature-2025} showed impressive throughput of $\sim$2GB/s as in the case of ONIX \cite{newman-multimodalneuralsystem-nature-2025}.
Furthermore, ONIX is an Open Neuro Interface-compliant system with open-source custom-built hardware for neuroscience experiments.
Both systems focus on single-host acquisition and are limited to Linux.

DelPreto et al. \cite{delpreto-actionsense-neurips-2022} addressed the specific use case in synchronized multi-host physiological recording with the open-source ActionSense.
To the best of our knowledge, this was the most lightweight and developer-friendly framework for multimodal physiological sensing.
It has a major limitation in throughput for multiple high-resolution videos, in reliance on retrospective data synchronization after a collection, Tab.~\ref{tab:challenges}(5), and in collation of modalities in post-processing from data recorded externally to it by 3rd party software.

ROS2 has played a pivotal role to system integration and engineering productivity, powering the new stage of robotics revolution \cite{macenski-ros2-science-2022}.
It is an extensive ecosystem for event-based distributed communication, that runs on one of the many implementations \cite{bode-dds-middleware-2023} of the Data Distribution Service (DDS) standard \cite{omg-dds-2015}.
It offers numerous integrations, toolboxes, simulators for design, simulation and validation of platform-independent publish-subscribe data-distribution systems.
To circumvent the challenge of using it on embedded devices, micro-ROS \cite{belsare-microros-springer-2023} was developed to transparently integrate with the rest of the ecosystem.
ROS2 has a high barrier to entry for new users \cite{daubaris-ros2challenges-rose-2023}, shows limited evidence of the use of realtime theory for scheduling of time-critical controls in robotic applications \cite{blass-ros2latency-rtas-2021}, degrades in performance under the distributed setting \cite{albatati-ros2survey-preprints-2024}, Tab.~\ref{tab:challenges}(10), and is ambiguous in the capacity for multiple high-resolution video streams, Tab.~\ref{tab:challenges}(11).
ROS2 is scoped to communication in robotics subsystems, and does not focus on precise synchronization across data streams \cite{elmakrini-physiosense-ram-2025}, Tab.~\ref{tab:challenges}(7).
It leverages community-driven implementations to support in-the-loop processing Tab.~\ref{tab:challenges}(8) and visualization Tab.~\ref{tab:challenges}(4).
The DDS-compliant middleware powering ROS2 was shown to have higher latency, lower throughput, and higher resource utilization than alternative middleware \cite{kang-middleware-m4iot-2020, ubicoders-middleware-youtube-2025}.

A plethora of open- and closed-source systems were built to address particular needs of researchers for data collection and stream processing, of new biomarkers and sensor technologies \cite{roddiger-openearable-imwu-2025, matsumura-edgepatch-devices-2025, moin-wearablepatch-nature-2021}, on dedicated hardware \cite{ma-nymeria-eccv-2024, lepold-iswc-harnode-2025, wiese-multimodaliot-arxiv-2025, ma-nymeria-eccv-2024}, or on common computing devices \cite{strandquist-athomecollection-jove-2023, chromik-sensorhubapp-sensors-2022, blum-smartphonelsldaq-sensors-2025}.
These dedicated hardware works are out of the scope of the present study, which focuses on a unified software framework that harmonizes fragmented hardware.

%% file: tables/system_requirements.tex
\begin{table*}[t]
  \caption{Requirements matrix for out-of-the-box continuous realtime multimodal physiological sensing and AI processing.}
  \label{tab:challenges}
  \begin{minipage}{\linewidth}
  \begin{center}
  \begin{tabular}{l c c c c c c c c c c c c c c c}
    \toprule
    \multicolumn{1}{l}{\textbf{Work}}
    & \multicolumn{14}{c}{\textbf{Criteria}}
    \\ \cmidrule{2-16}
    & \rotatebox[origin=c]{90}{\makecell{Opensource}}
    & \rotatebox[origin=c]{90}{\makecell{Multimodal}}
    & \rotatebox[origin=c]{90}{\makecell{Synchronized}}
    & \rotatebox[origin=c]{90}{\makecell{Custom AI\\in-the-loop}}
    & \rotatebox[origin=c]{90}{\makecell{Extendible w/\\custom sensors}}
    & \rotatebox[origin=c]{90}{\makecell{Crash safe}}
    & \rotatebox[origin=c]{90}{\makecell{Data\\collection}}
    & \rotatebox[origin=c]{90}{\makecell{Stream\\processing}}
    & \rotatebox[origin=c]{90}{\makecell{Fixed lab\\suitable}}
    & \rotatebox[origin=c]{90}{\makecell{Free-living\\suitable}}
    & \rotatebox[origin=c]{90}{\makecell{Distributed\\multi-host}}
    & \rotatebox[origin=c]{90}{\makecell{Handles\\high-res video}}
    & \rotatebox[origin=c]{90}{\makecell{Platform\\agnostic}}
    & \rotatebox[origin=c]{90}{\makecell{Available in\\Python}}
    & \rotatebox[origin=c]{90}{\makecell{Visualization}}
    \\
    \midrule
       LabView \cite{elliott-labview-jala-2007}
       &
       & \checkmark
       &
       &
       & \checkmark
       & \NA
       &
       & \checkmark
       & \checkmark
       & \checkmark
       &
       &
       & \NA
       & \NA
       &
       \\
       Vicon \cite{vicon}
       &
       & \checkmark
       & \checkmark
       &
       &
       &
       & \checkmark
       &
       & \checkmark
       &
       &
       & \checkmark
       & 
       & 
       & \checkmark
       \\
       MVN Analyze \cite{xsens}
       &
       & \checkmark
       & \checkmark
       &
       &
       &
       & \checkmark
       &
       & \checkmark
       &
       &
       &
       &
       &
       & \checkmark
       \\
       HARNode \cite{lepold-iswc-harnode-2025}$^1$
       & \checkmark
       &
       & \checkmark
       &
       & \checkmark
       & ?
       & \checkmark
       &
       & \checkmark
       & \checkmark
       &
       &
       & \NA
       &
       &
       \\
       MQTT protocol
       & \checkmark
       & \checkmark
       & \NA
       & \NA
       & \checkmark
       & \NA
       & \NA
       & \NA
       & \checkmark
       & \checkmark
       & \checkmark
       &
       & \checkmark
       & \checkmark
       & \NA
       \\
       MQ Systems \cite{fu-mqsystems-ieee-2021}
       & \checkmark
       & \checkmark
       & \NA
       & \NA
       & \checkmark
       & \checkmark
       & \checkmark
       & \checkmark
       & \checkmark
       & \checkmark
       & \checkmark
       & 
       & \checkmark
       & \checkmark
       &
       \\
       NATS \cite{nats}
       & \checkmark
       & \checkmark
       & \NA
       & \NA
       & \checkmark
       & \checkmark
       & \checkmark
       & \checkmark
       & \checkmark
       & \checkmark
       & \checkmark
       & 
       & \checkmark
       & \checkmark
       &
       \\
       Syntalos \cite{klump-syntalos-nature-2025}$^2$
       & \checkmark
       & \checkmark
       & \checkmark
       &
       & \checkmark
       &
       & \checkmark
       & \checkmark
       & \checkmark
       &
       &
       & \checkmark
       & 
       & \checkmark
       & \checkmark
       \\
       ONIX \cite{newman-multimodalneuralsystem-nature-2025}$^2$
       & \checkmark
       & \checkmark
       & \checkmark
       &
       & \checkmark
       & \checkmark
       & \checkmark
       & \checkmark
       & \checkmark
       &
       & \checkmark
       &
       & \checkmark
       & \checkmark
       \\
       LSL \cite{kothe-lsl-biorxiv-2024}
       & \checkmark
       & \checkmark
       & \checkmark
       &
       & \checkmark
       &
       & \checkmark
       & \checkmark
       & \checkmark
       & \checkmark
       & \checkmark
       & \checkmark$^3$
       & \checkmark
       & \checkmark
       & \checkmark$^4$
       \\
       ActionSense \cite{delpreto-actionsense-neurips-2022}
       & \checkmark
       & \checkmark
       & \checkmark$^5$
       &
       & \checkmark
       &
       & \checkmark
       &
       & \checkmark
       & \checkmark
       &
       &
       & \checkmark
       & \checkmark
       & \checkmark
       \\
       PhysioSense \cite{elmakrini-physiosense-ram-2025}
       & \checkmark
       & \checkmark
       & \checkmark
       &
       & \checkmark
       & \checkmark
       & \checkmark
       &
       & \checkmark
       & \checkmark
       &
       & \checkmark$^6$
       &
       &
       & \checkmark
       \\
       ROS2 \cite{macenski-ros2-science-2022}
       & \checkmark
       & \checkmark
       & \checkmark$^7$
       & \checkmark$^8$
       & \checkmark
       & \checkmark
       & \checkmark$^9$
       & \checkmark
       & \checkmark
       & \checkmark
       & \checkmark$^{10}$
       & \checkmark$^{11}$
       & \checkmark
       & \checkmark
       & \checkmark$^4$
       \\
       CLAID \cite{langer-claid-fgcs-2024}$^{12}$
       & \checkmark
       & \checkmark
       & \checkmark
       & \checkmark
       & \checkmark
       & ?
       & \checkmark
       & \checkmark
       & \checkmark
       & \checkmark
       &
       & \checkmark$^6$
       & \checkmark
       & \checkmark
       & \checkmark
       \\
       \textbf{HERMES [ours]}
       & \checkmark
       & \checkmark
       & \checkmark
       & \checkmark
       & \checkmark
       & \checkmark
       & \checkmark
       & \checkmark
       & \checkmark
       & \checkmark
       & \checkmark
       & \checkmark
       & \checkmark
       & \checkmark
       & \checkmark
       \\
    \bottomrule
  \end{tabular}
  \end{center}
  \footnotesize{\emph{Note:} "\NA" is "Not applicable", "?" is unknown or ambiguous.}
  \end{minipage}
\end{table*}

%% file: content/03-methodology.tex
\subsection{Continuous Realtime Edge AI Methodology}
\label{sec:methodology}
Multimodal data acquisition in the real-world is inherently challenging and inevitably prone to missing data \cite{xu-lsm-arxiv-2025}.
Old methods relied on imputation, data filtering, synthesis, and discarding, limiting the overall effectiveness of deployed models because of the discrepancy between the cleaned training data and the real-world \cite{ding-biosignalllm-arxiv-2024}.
This study considers methods that directly learn from the incomplete data \cite{xu-lsm-arxiv-2025} or naturally account for it \cite{narayanswamy-foundationwearablemodels-arxiv-2024, ding-biosignalllm-arxiv-2024, zhu-imudatadrop-jbhi-2024} without introducing imputation bias and dropping stretches with missing data, more principled and generalizable to the real-world.

To overcome the challenges of the real-world, we propose in the following sections a methodology that aligns with the nature of physiological sensing and realtime AI processing in the wild.
Online inference is generalized to two practical extremes, which contextualizes fusion of heterogeneous modalities in the presence of missingness, and offers a pragmatic outlook on synchronization.
This context fosters a system-level perspective on deployed systems, addressing fragmentation between offline data collection and online processing in the real-world.
The holistic approach guides downstream AI model development, offering trade-offs to balance between requirements and apparent application constraints.

\input{figures/methodology}

\subsubsection{Definitions}
\label{sec:definitions}
This work defines "multimodal" as streams of data from the same or different host devices, that are generated by sensor(s) asynchronously, Fig.~\ref{fig:methodology}a.
A setup of four external cameras, each on its own physical interface, has four video modalities.
A motion capture system that sends interleaved UDP packets of bundled measurements - a bundle of 3D joint coordinates, a bundle of joint angles, a bundle of inertial data, has three modalities.
A pair of eye-tracking smartglasses streaming egocentric vision of 30 frames-per-second (fps) and gaze vectors at a variable rate of 0-120 Hz, has two modalities.
Downstream AI processing components are not impacted by this upstream convention and are flexible to redefine "multimodal" according to user needs, e.g. treating bundled accelerometer and gyroscope data as separate modalities.

In contrast to the fragmented meaning of "realtime" in robotics and operating systems, we treat it as the ability of a continuously running system to complete all its processing at the same rate as the rate of arrival of new measurements, Fig.~\ref{fig:methodology}b.
This definition aligns to the objective of online sensing and inference systems, and is comprehensively validated with benchtop stress testing.

The vital concept of "synchronization" is frequently underexplored in online distributed sensing and data collection literature.
Synchronicity across distributed signals can only exist within tolerances - milli-, micro-, nanosecond, depending which level is considered acceptable by the use case.
The present work considers signal synchronicity as a tight continuous agreement in the notion of time across multiple distributed sensing devices - the lowest skew between device clocks at any instance in time that allows to directly draw positive pairs of samples across modalities.

\subsubsection{Realtime Multimodal Inference Strategies}
\label{sec:inference_strategies}
To align with the asynchronous and heterogeneous nature of the underlying multimodal signals \cite{wu-asyncmultimodalinference-sensys-2024}, two inference strategies are proposed in Fig.~\ref{fig:methodology}c.
The \texttt{PUSH} strategy triggers inference in an event-driven way, on every new measurement arrival from either modality. 
The method relaxes data alignment requirements and operates in a low-latency mode, albeit limited to smaller AI models due to the higher frequency of predictions, hence shorter processing time budget. 
The \texttt{PULL} strategy performs computations sparsely or at a user-specified rate.
It buffers independent streams, to adjust for temporal skews due to the different propagation and sensing delays, and draws positive pairs of temporally correlated samples.

The former is flexible to drive inference as soon as any of the asynchronous streams supplied a new measurement, but leaves ambiguity on how to merge different modalities.
The latter incurs worst case latency up to the longest propagation delay of a stream.
That makes fusion of data unambiguous to draw positive pairs of data by the closest acquisition time across modalities, but requires robustness toward missing samples.
Domain expertise-driven combination of the two approaches is a valid option that balances between the two extremes.

Multivariate time-series data contains hidden patterns that describe the temporal dynamics of the sensed phenomenon.
Prior works explore ways to most effectively extract these insights from overlapping windows of temporal data.
Orthogonal to the inference strategy, Fig.~\ref{fig:receptive_field} shows that selection of the window size (receptive field) and the size of overlap between consecutive windows, influences the computational intensity and the processing time budget, respectively \cite{yudayev-rtstgcn-mlsp-2024}.
We propose a coupled approach of varying these factors along with the inference strategy.
Increasing \textit{s} k-times boosts the time budget by k, supporting more computationally intensive models with a larger \textit{L}, at the expense of k-times sparser predictions.
This approach contributes a pragmatic additional degree of freedom to fit the constraints by the downstream AI model, if the end use case requirements are tolerant to reduced temporal resolution of predictions and any architectural model changes would cause significant AI performance degradation.

\input{figures/receptive_field}

To aggregate continuously captured multimodal data for downstream AI processing, it is required to address the heterogeneity of sample rates, dimensionality, and missingness in the data.
State-of-the-art methods \cite{xu-lsm-arxiv-2025, zhang-sensorlm-arxiv-2025, wang-multimodalfusion-informationfusion-2025} embed each modality into a unified intermediate temporal embedding, generalizing to the continual and missing nature of signals in the real-world.
To generically accommodate different use cases, we propose a decoupled approach to data exchange between sensing and AI processing components that enables strategies selection at the data consumer end.

\subsubsection{Continuous Synchronization}
\label{sec:continuous_synchronization}
Fusion of distributed streams into a common representation relies on the temporal alignment of the contained data \cite{ni-multimodalwearablesurvey-arxiv-2024}.
When data is captured by a single integrated device like \cite{zhang-sensorlm-arxiv-2025, xu-lsm-arxiv-2025}, alignment is intrinsic with respect to the host.
In an environment with multiple sensing hosts and distinct sensors, interconnected in a wireless or wired network, drawing positive pairs of samples across distributed modalities becomes a challenge \cite{li-multimodalsurvey-arxiv-2024}.
Distributed sensors do not have a common notion of time, and temporal misalignment of data has detrimental effects on downstream AI performance \cite{wolling-timediscrepancy-iswc-2025, trollmann-timediscrepancy-thesis-2025, klump-syntalos-nature-2025}.
Building dedicated synchronization hardware similar to Meta's \cite{ma-nymeria-eccv-2024} is out of reach for many researchers, and has limited sensor integration support.

We propose to align distributed modalities by continuous synchronization of logical clocks across host devices.
Physical clocks are divergent monotonic piezo oscillators with equal length periods between oscillations, experiencing infinitesimal pico- or femtosecond jitter, and influenced by temperature and applied voltage \cite{kleppmann-concdissys-cam-2021}.
Inherent mechanical differences between oscillators of distributed devices result in minor drift that compounds overtime into substantial clock differences \cite{wolling-timediscrepancy-iswc-2025}.
Depending on the category of the device, time-keeping functionality varies \cite{wu-multimodalsensingdemo-sensys-2025}.
Sophisticated hosts have built-in network-based synchronization, while most sensors are free-running and synchronization is rarely addressed.
Jitter in a synchronized device's time, with respect to the reference clock, characterizes the consistency of continuous synchronization.
Consensus on the synchronization tolerance is largely understudied in the literature \cite{wolling-timediscrepancy-iswc-2025}.
We consider the level of synchronicity widely acceptable $\forall t_{skew,i\rightarrow j} \leq \frac{1}{2}t_{sample}$, where $i$ and $j$ distinct devices, and $t_{sample}$ the highest sample rate, Fig.~\ref{fig:methodology}d. This level of host synchronization satisfies the requirement for unambiguous cross-modality sample pairing.

For the current methodology to be broadly applicable to commercial and custom sensors, we generalize distributed synchronization process into host-to-host, and sensor-to-host categories.
Host-to-host synchronization leverages existing local network time-keeping facilities like PTP and NTP.
Sensor-to-host connections with limited built-in synchronization functionality use roundtrip time measurements to estimate the propagation delay of a sensor, with respect to its host.
The estimate is used as a relative offset from the time-of-arrival of the samples at the host, to extract true sampling time.
This indirect approach, visualized in Fig.~\ref{fig:methodology}d, leverages highly accurate synchronization between distributed hosts to align samples across locally and remotely captured modalities.
Performing the procedure periodically ensures synchronicity system-wide, in the presence of non-deterministic external factors like temperature-related drift of oscillators and varying propagation delays of wireless sensors.

%% file: figures/methodology.tex
\begin{figure*}[t]
  \centering
  \includegraphics[width=\linewidth]{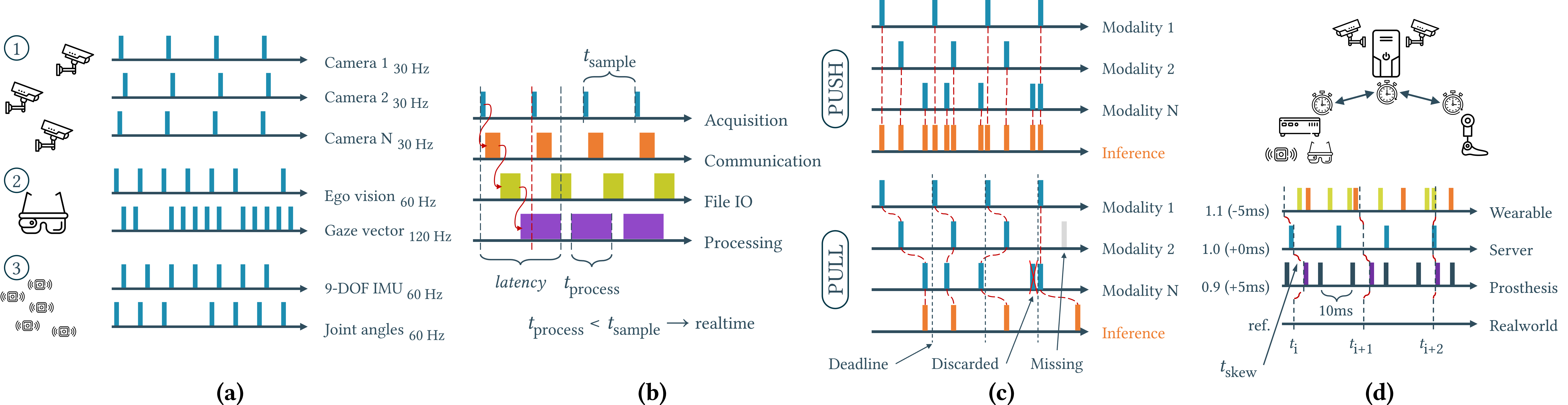}
  \caption{Generalization of multiple modalities (a): same data stream from multiple identical sensors (a.1), different sample rate data streams of the same sensor (a.2), separate same sample rate stream per bundle of measurements of the same sensor (a.3).
  Definition of realtime operation and latency in pipelined data flow (b).
  Two proposed extremes of continuous inference strategies: asynchronous event-driven (push) on any new modality update, and buffered alignment (pull) for the latest samples across modalities (c).
  Synchronization of distributed networked devices and connected sensors (d).}
  \Description{
  In "a", arrival of packets over time from different types of sensors, generalizing multimodality in different functioning sensors into common categories: three cameras streaming at a consistent and reliable 30 frames-per-second with a slight skew from one another due to propagation delay difference, a pair of smartglasses streaming with occasional drops packets of gaze vector data at 120 Hz and egocentric video at 60 frames-per-second, an inertial motion capture system streaming at 60 Hz in separate packets bundles of joint angles and raw kinematics data of all individual sensors.
  In "b", multiple tasks - acquisition, communication, file input-output operations, processing, forming a data processing pipeline, passing data from one to another.
  The cumulative elapsed time between acquisition event and end of processing is shown as latency. The system is shown as realtime if each of the tasks is consistently completed at the same rate as the data acquisition rate between consecutive samples.
  In "c", push and pull strategies extremes for multimodal alignment.
  Push triggers inference on each newly arrived sample across any modality, showing tightly packed event-driven inferences over time, potentially with little difference in environment in between.
  Pull shows a conditional inference that triggers on-demand or at predefined time increments, either when all modalities received the most recent valid data, or when the deadline to make a prediction has been reached - even if some modalities have not captured a new sample.
  In "d", local notions of time of three distinct hosts - edge server, prosthesis, wearable minicomputer, with respect to the real-world notion of time.
  The hosts exchange local clock information over the local network with each other, to agree on the common time.
  Skew between clocks at the start of the displayed period, erroneously correlates samples of video from the edge server, to a telemetry measurement on the prosthesis that is one sample later than the sample that corresponds to the closest instance as when the video frame was captured.
  Over time, the hosts are shown to minimize the clock skew and able to correlate samples without ambiguity.}
  \label{fig:methodology}
\end{figure*}

%% file: figures/receptive_field.tex
\begin{figure}[h]
  \centering
  \includegraphics[width=0.9\linewidth]{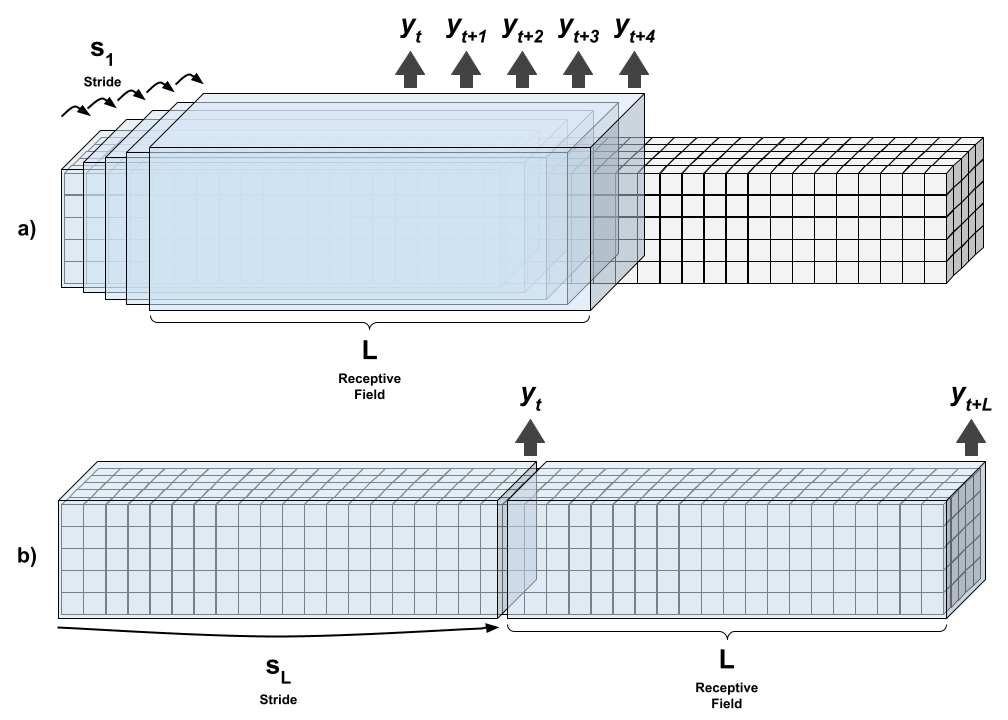}
  \caption{Relationship between the AI receptive field and the overlap of consecutive windows. (Adapted from \cite{yudayev-rtstgcn-mlsp-2024}.)}
  \Description{
  XXXXXXXXXXXXXXXXX
  }
  \label{fig:receptive_field}
\end{figure}

%% file: content/04-architecture.tex
\subsection{Sensing and Inference System Architecture}
\label{sec:system_architecture}
To accelerate research in real-world physiological edge AI systems, we release HERMES - an open-source developer-friendly Python framework for realtime multimodal sensing and processing at the heterogeneous distributed edge.
It fills the gaps in the research domain by providing a testbed for novel physiological edge AI methods validation, closed-loop clinical interventions studies, curation of high-quality multimodal datasets, and for building patient-centered intelligent health applications.

This modular hardware-agnostic framework is built to embody the methodology described in Sec.~\ref{sec:methodology}, and is available on all major operating systems.
It generalizes internal components to streamline interfacing to any commercial, open-source, or custom sensors and actuators.
Integration of a device into the unified streaming acquisition system is realized by wrapping the device's official software development kit (SDK) or standard wired and wireless communication protocols, like Bluetooth, TCP/UDP, USB, ANT, MQTT, etc., into templates provided by \verb|HERMES|.
Implementation of many popular commercial physiological sensors is available in \verb|HERMES| out-of-the-box, and the framework is periodically extended with support for new devices.
A PyTorch integration equips sensing applications with closed-loop AI processing functionality of models provided by the user.

\verb|HERMES| is built on top of the fast and cross-platform ZeroMQ transport \cite{hintjens-zeromq-2013}, FFmpeg-accelerated video processing \cite{ffmpeg}, memory efficient write-optimized datastructures, parallelism and asynchronous cooperative multitasking, and data file formats suited for conventional AI training workflows.
It provides low-latency high-performance event-driven multimodal synchronized sensing and computation, crucial for the expanding field of intelligent healthcare edge applications.
Benchtop characterization of the final sensing system with user's selected devices contextualizes the target application and guides downstream AI model design with well-defined operational constraints.
The following sections elaborate on the technical aspects that facilitate the high-performance system architecture of the framework.

\input{figures/design}

\subsubsection{Layers of Abstraction}
\label{sec:abstraction}
The complexity of a distributed heterogeneous system is masked behind the abstraction model of Fig.~\ref{fig:design}a.
By decoupling the logical flow of data from the hardware's physical layout, user's processing logic gains transparent access to the data of interest that resides elsewhere.
This way, a prosthesis embedded controller receives user movement intent predictions, computed by the dedicated AI wearable single-board computer on information from sensors connected to it, and those embedded into the prosthesis, Fig.~\ref{fig:teaser}.

The physical connections layer establishes the backbone network between the host devices, and connects to the corresponding sensors and actuators - sources and sinks, respectively.
At the device interface layer, each host is responsible for wrapping its local sources and sinks into a generalized event-driven \texttt{Node} that interacts with the device via an SDK, a standard protocol, or an operating system socket, passing new data to the messaging layer with a unique \textit{topic} descriptor, Fig.~\ref{fig:design}b.

The abstraction model generalizes any sensing, actuation, or processing block into a \texttt{Node} that handles core message exchange. 
Each node is an instance of \texttt{Producer}, \texttt{Consumer}, or \texttt{Pipeline}, that only generates, only receives, or processes incoming data to generate new from it, respectively.
Figure~\ref{fig:design}b expands on the overview of Fig.~\ref{fig:teaser} and offers an internal look onto the main building blocks present at each host.
The message exchange layer, powered by the ZeroMQ high-throughput low-latency core middleware \cite{lacorte-messagingperformance-arxiv-2025}, establishes persistent inter- and intra-host communication links - between the networked hosts, and between a host's local nodes.
Data submitted to the layer is then intelligently routed via these links in a PUB/SUB pattern \cite{vansteen-distributedsystems-online=2023} to any node interested in the data corresponding to this unique \textit{topic}, Fig.~\ref{fig:design}c.
The host's \texttt{Broker} block manages the routing of information and the lifecycle of all the nodes of the corresponding host.

Each node defines the list of topics it is interested in, performs user-defined logic with that data, and optionally publishes data back to the message exchange layer with a new descriptor.
Because ZeroMQ is a platform-independent transport, with bindings to all popular programming languages, migrating existing sensors, actuators or algorithms from another framework to \verb|HERMES| is achievable by injecting the matching ZeroMQ messaging operation, Alg.~\ref{alg:pseudocode} (\textit{line} \ref{lst:pseudocode:publish}), in the user's existing code, Alg.~\ref{alg:pseudocode} (\textit{line} \ref{lst:pseudocode:loop}).
In practice, a user is only required to extend the base \texttt{Producer}, \texttt{Consumer}, or \texttt{Pipeline} with custom processing logic to abstract away from the complexity of distributed communication, and to ensure the added block harmoniously works with the rest of the system.

\input{algorithms/pseudocode}

\subsubsection{Streaming High-Throughput Datastructure}
\label{sec:datastructure}
To enable high-throughput streaming of data, an efficient and thread-safe datastructure is required.
Dynamic arrays are contiguously stored blocks of allocated system memory that have a low theoretical computational complexity and constant read time.
In practical applications of streaming, with no predefined length limit and large elements (e.g. images), dynamically-sized arrays have a high cost for write operations because they require reallocation by the operating system to accommodate a bigger contiguous block of memory.
The operation takes a big toll on the CPU, is slow, and causes out-of-memory issues in high-intensity applications, making them impractical for the low-latency streaming requirements.

\input{figures/design_internal}

\verb|HERMES| interacts with data via the \texttt{Stream} datastructure, Fig.~\ref{fig:design_internal}a.
It is based on the doubly-linked list that dynamically allocates memory for each individual streamed sample, without storing data contiguously.
This yields a characteristic constant access time for operations at either end of the datastructure, but access to a random element in the datastructure has an \textit{O(n)} complexity.
Reading a random element in a doubly-linked list requires traversing the datastructure until the element is found.
By system design, we only access the datastructure like a conveyor belt: the head to insert the newest sample, and the tail to pop the oldest.
This enables simultaneous reading and writing of the datastructure at each end, making it ideal for the low-latency high-throughput streaming context.

\subsubsection{Video Data Processing}
\label{sec:video_data_processing}
A generalized camera acquires frames continuously and transfers each directly into host's reserved kernel memory area via direct memory access (DMA), without the interaction of the CPU, Fig. \ref{fig:design_internal}b.
Multiple cameras are connected to a peripheral input-output (IO) device of the host - e.g. the network interface card (NIC) for GigE vision cameras.
The IO device writes interleaved frames by DMA into a fixed-length shared circular buffer, e.g. of up to 12 frames, in a round-robin fashion.
If frames in the circular buffer are not consumed fast enough by the user application, the oldest frames are overwritten.

We ensure that no frames written via DMA into memory are lost if user's custom logic is not fast enough.
\texttt{HERMES} moves each newly available frame from the kernel space circular buffer into the corresponding \texttt{Stream} in the safe, larger, user memory space of the system.
That data is then available through the datastructure interface to other blocks, Sec.~\ref{sec:datastructure}, like the \texttt{Storage} component that safely and efficiently stores video to disk.

\subsubsection{Persistent Data Storing}
\label{sec:persistent_data_storing}
Storing streaming data is the responsibility of the dedicated asynchronous \texttt{Storage} thread inside each host, Fig.~\ref{fig:design_internal}a, which flushes accumulated data from the \texttt{Stream} datastructures to disk.
File IO operations are slow, but many files can be written by the processor concurrently.
\texttt{Storage} exploits this characteristic to facilitate efficient flushing of acquired realtime physiological data to disk.
This keeps memory usage in check and protects it from overflowing, for a resilient system.
Such a system runs and collects data indefinitely, unless power is cut, or the device runs out of disk space: both of which are the responsibility of the user.
The system flushes data periodically, according to the tunable \textit{period} parameter, which prevents loss or corruption of data in case of a host crash: no more data is lost than the selected period.
The parameter influences the rate at which memory accumulates and CPU is utilized, giving user flexibility to balance between worst case reliability, desired memory footprint and CPU utilization.

The sawtooth wave in Fig.~\ref{fig:design_internal}c shows the stable steady-state memory footprint of a high-throughput system.
Memory accumulates as sensors generate new data into \texttt{Streams}, and clears when the \texttt{Storage} component activates to flush available data to disk for persistent storage.
Similar to video data safe-keeping method of Sec.~\ref{sec:video_data_processing}, other modalities are safely captured: a fast dedicated thread moves newly available data from the interface layer of the related device to its corresponding \texttt{Stream}.
Once data is in user memory space, it is managed by other components and user's custom logic.
This approach ensures that all data that enters the hardware is robustly retained by our software and never lost.

\input{figures/synchronization_details}

\paragraph{Video Data}
\label{sec:video_data}
High-resolution videos require large space for storage.
A single raw 30 fps 1440p video with 8-bit colors generates 332 MB/s of data.
Storing or transferring such volume of data in the raw pixel format is not practical.
Encoding decreases the required space while preserving perceptually important information by intelligently compressing repetitive areas in a frame (intra-frame) and repetitive areas in consecutive frames of a video (inter-frame) \cite{sze-hevc-springer-2014}.

Realtime encoding is challenging and \texttt{HERMES} integrates with the mature hardware-accelerated video processing ecosystem - FFmpeg \cite{ffmpeg}.
We continuously pipes captured high-resolution live video by inter-process communication, frame-by-frame, into an FFmpeg subprocess, one per video.
Integration with FFmpeg enables our system to efficiently process realtime videos by leveraging dedicated high-performance hardware encoders available to the host device - integrated in the processor or via an external accelerator, like a graphics card.
Several tuned presets for the hardware platforms used in our study are distributed for out-of-the-box writing of ubiquitous MP4/MKV files with H.264/HEVC codecs.
The presets are tuned through a configuration file, for user convenience, which are accessed at system runtime.

\paragraph{HDF5 Data}
\label{sec:hdf5_data}
All non-video data has predefined data type and dimensionality, and is treated as streamlined N-dimensional arrays.
Large volume of time-series data from longitudinal studies and long data collections requires researchers to leverage a server or a high-performance computer (HPC) for downstream AI training, before deploying a model to the edge.
In this processing paradigm, a robust parallel file format is required for efficient use of resources in existing AI training workflows.
In the HPC, AI, and scientific computing context, HDF5 \cite{folk-hdf5-ad-2011} stands out as the most widely used format across practitioners.
For smoother integration into existing workflows, \texttt{HERMES} adopts the same efficient file format.

\subsubsection{Synchronization of Modalities}
\label{sec:synchronization_of_modalities}
To draw correct positive pairs across distributed modalities, the framework timestamps all data samples captured by a host, using its high-precision system clock, as soon as the samples from its locally connected sensors arrive at an interface, Fig.~\ref{fig:synchronization_details}a.
This time-of-arrival reduces sensor-to-host synchronization challenge to a simple propagation delay estimation to adjust the time-of-arrival, to extract true time-of-generation.

All distributed hosts in the sensing setup are kept in-sync by the local NTP/PTP time server.
It maintains a common notion of time across the interconnected hosts using their sophisticated time-keeping functionality.
Each locally connected sensor stream is principled to this trusted notion of time by its respective host using the time-of-arrival.
Any drift between the stated and the actual sample rate of a sensor, due to mechanical oscillator differences, is mediated.
It is not critical for synchronization of streams that sensors have integer sample rate with respect to a reference clock and each other, i.e. without any drift.
It is crucial that regardless of the actual sample rate of the sensors, samples from different streams are related to a common reference time, to heuristically draw positive pairs by pairing closest samples across modalities.

\subsubsection{Transmission Delay Realignment}
\label{sec:transmission_delay_realignment}
Continuous realignment of streaming data is needed for two reasons.
One, to couple samples from individual asynchronous sensors within the same modality that correspond to the same time instance - e.g. multiple wireless IMUs with an internal wireless continuous clock synchronization mechanism, asynchronously sending data to the host \cite{alcala-dots-movella-2021}.
The synchronized devices transmit their respective measurements during each time window, in an overlapped fashion.
Congestion on the shared wireless medium has non-deterministic influence on the order and the delay of the asynchronously received packets.
Even though individual samples were captured simultaneously by the coordinated sensors, they arrive to the host out-of-order, and must be realigned for reliable use.
Two, to counteract the impact of the variable transmission or internal delay of a stream in the \texttt{PULL} strategy, prior to drawing positive pairs of samples across modalities for fusion.

Our system offers a buffered continuous alignment method with a tunable \textit{stale} parameter for content-based alignment of case one, Fig.~\ref{fig:synchronization_details}b, and with a relative offset into each stream to align modalities by the extracted true time-of-generation of case two.
The \textit{stale} parameter specifies the maximum tolerance to delays across intra-modality sensors.
Once the specified value has been exceeded, the lagging sample is considered lost and the buffer yields incomplete data downstream.
Tuning the parameter balances additional latency and tolerance to late arrivals.
The relative offset is the periodically measured roundtrip time to the sensor for adjusting the time-of-arrival used to align modalities.
It is provided per modality, is used to retrieve positive pairs of sample on each \texttt{PULL}, and is updated on each roundtrip time estimation.

In contrast to the fixed internal alignment across modalities (case two) of other frameworks \cite{wang-lslscope-springer-2023}, \texttt{HERMES} brokers data with the lowest latency possible in an event-driven way, and provides periodically measured roundtrip time along with each modality.
Downstream data consuming nodes are expected to choose the desired alignment strategy to locally align the modalities and balance between added latency and precise alignment.
Combined with Sec.~\ref{sec:synchronization_of_modalities}, modalities can be reliably fused.

%% file: figures/design.tex
\begin{figure*}[t]
  \centering
  \includegraphics[width=\linewidth]{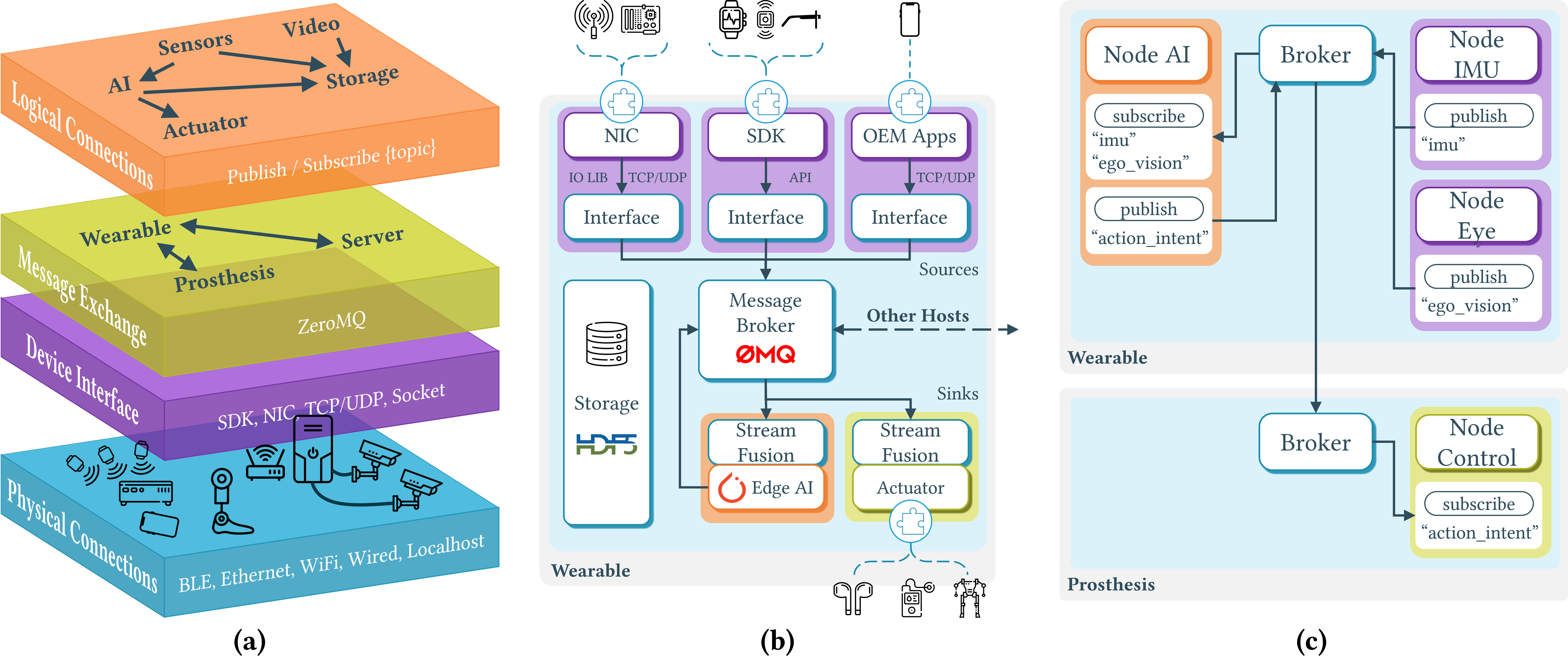}
  \caption{Abstraction model that makes access of logic blocks to data of interest transparent across the network and devices, regardless of data locality (a).
  Decoupled system components at each host (b) - publisher (purple), consumer (green), pipeline (orange), fusion blocks responsible for data alignment and embedding strategies, storage component captures all sensed and processed data.
  Flow of message exchange based on \textit{topic} data descriptors through a \texttt{Broker} (c).}
  \Description{
  In "a", logical layers model of the system in the bottom-up order.
  Physical connections layer between sensors and hosts using wired and wireless interfaces like bluetooth, WiFi, Ethernet, proprietary protocols, and localhost for communication with locally hosted software that produces data.
  Device-specific interface layer to control the device and received data from it, like the corresponding software development kit, 3rd-party software application, operating system socket, network interface card.
  Message exchange layer for persistent network communication link between host devices.
  Logical connections between functional Node modules performing predetermined function and having transparent access to any modality of data across the distributed setup, exchanging it in a publish-subscribe manner regardless of its physical origin.
  In "b", a detailed structure of a Node.
  Sensing devices connect physically to the corresponding host as sources of data, the host interfaces with each according to the available or implemented interface and wraps that logic into an implementation of an abstract Producer class.
  The sources transmit every new packet to the message broker thread that intelligently routes the data locally and to external hosts if any are interested in its data.
  The host has some sink nodes, an AI processing node implementing the abstract Pipeline class, and a device actuating node that implements the abstract Consumer class to mask the complexity of interacting with an external intervention device like an earpiece or a motor.
  In "c", flow of data between remote hosts in a publish-subscribe communication pattern, based on the "topic" descriptor of the associated data.
  Only data that was requested by a node or a host is actually exchanged on the network.}
  \label{fig:design}
\end{figure*}

%% file: algorithms/pseudocode.tex
\begin{algorithm}
  \caption{Pseudocode of a \texttt{Node} interfacing its sensor and publishing acquired samples to the message exchange layer.}
  \label{alg:pseudocode}
  \begin{algorithmic}[1]
    \REQUIRE $\mathit{s} \in \mathit{S}$ \hfill \COMMENT{Unique "topic"}
    \WHILE{$true$} \label{lst:pseudocode:loop} 
      \STATE $D \gets GetNewestData()$ \hfill \COMMENT{User's sensor interface}
      \STATE $t \gets GetHostTime()$ \hfill \COMMENT{Time-of-arrival}
      \FORALL {$d \in D$}
        \STATE $Publish(s, d, t)$ \label{lst:pseudocode:publish} \hfill \COMMENT{Replace to migrate}
      \ENDFOR
    \ENDWHILE
  \end{algorithmic}
\end{algorithm}

%% file: figures/design_internal.tex
\begin{figure*}[t]
  \centering
  \includegraphics[width=\linewidth]{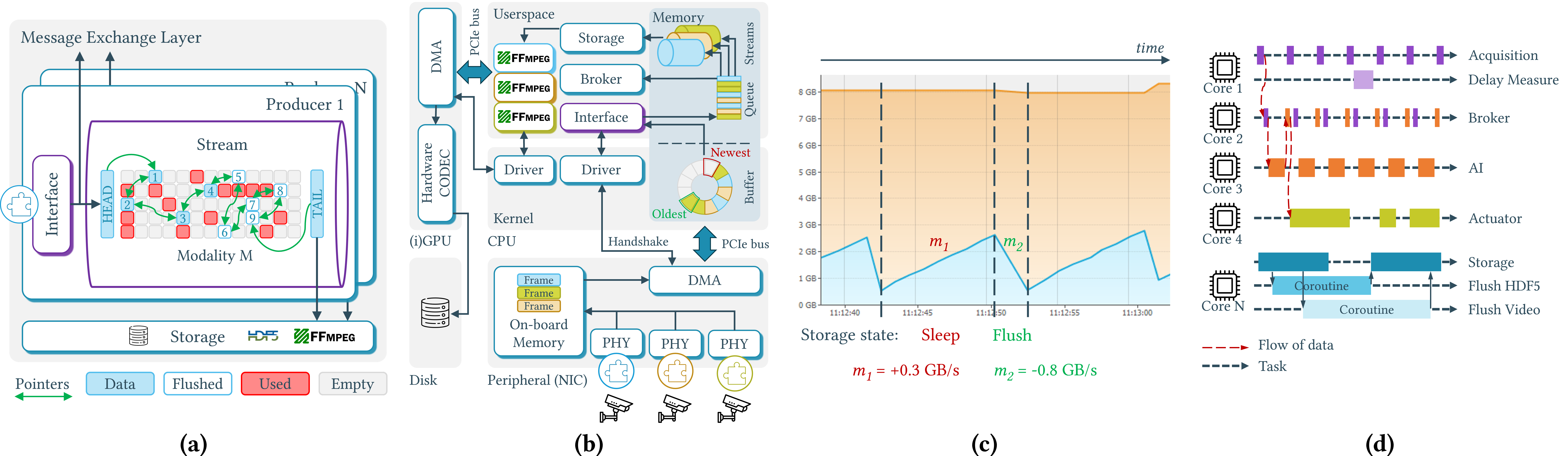}
  \caption{Efficient doubly-linked list FIFO datastructure for high-throughput low-latency streaming data. The \texttt{Producer} thread writes to it from the head and the \texttt{Storage} thread reads from it from the tail, simultaneously (a).
  Overview of the video data flow through the system (b).
  Memory pressure sawtooth profile - system keeps up with the streamed data if $\left|m_2\right| > \left|m_1\right|$: storage component sleeps during positive slope, flushes accumulated data during negative slope (c), adapted from \cite{ebner-memory-stack-2014}.
  Visualization of the CPU usage in the system implementation leveraging multiprocessing, multithreading, and asynchronous IO coroutines (d).}
  \Description{
  In "a", multiple producer nodes, each contains the stream datastructure. The distinct device interface captures a new sample from the corresponding sensor in an event-driven fashion and routes the sample to both, the host's message exchange layer and the corresponding datastructure. Once a sample is captured and memory is allocated for it, pointer to the element is obtained and added to the doubly-linked list-like structure without additional copies, updating the metadata head information. The storage module has access to the datastructures and flushes all available data, starting from the oldest, on-demand to disk for persistent storage as an HDF5 file, or through FFmpeg as MP4/MKV file formats for video data.
  In "b", flow of the video frames from the camera interface into the main message exchange system, and flushed to disk. The device interface connected to the cameras implements a driver that writes internally buffered frames for each connected device into the system's main memory via direct memory access, to offload CPU from peripheral tasks. The frames are written via direct memory access into the kernel memory space of the host in an interleaved manner in a fixed-size ring buffer that overwrites oldest frames with newest continuously even if data is not accessed by the user. The HERMES system moves interleaved frames in a first-in-first-out order into an intermediate queue in the user memory space to prevent video frames from being overwritten by the kernel driver. The frames are then deinterleaved and stored in the corresponding stream datastructure, as well as sent to the message exchange layer for any interested subscribers. When the storage module enters active flush periods, it flushes accumulated frames through the corresponding FFmpeg process to disk. The FFmpeg process corresponding to the video stream encodes video data using host's dedicated video codec accelerator into a mainstream compressed file format.
  In "c", a sawtooth-like chart of system memory pressure profile over time. A positive slope of 300 MBps during the inactive state of the storage module, and a negative slope of 800 MBps during the active flush state are displayed. The system maintains a stead state in memory use because it can flush data faster than it accumulates in memory.
  In "d", timeline of CPU usage across different processes of the system. The diagram emphasizes how the implementation uses multiprocessing, multithreading, and asynchronous IO coroutines, to construct a high-performance data acquisition and processing system.
  }
  \label{fig:design_internal}
\end{figure*}

%% file: figures/synchronization_details.tex
\begin{figure*}[t]
  \centering
  \includegraphics[width=\linewidth]{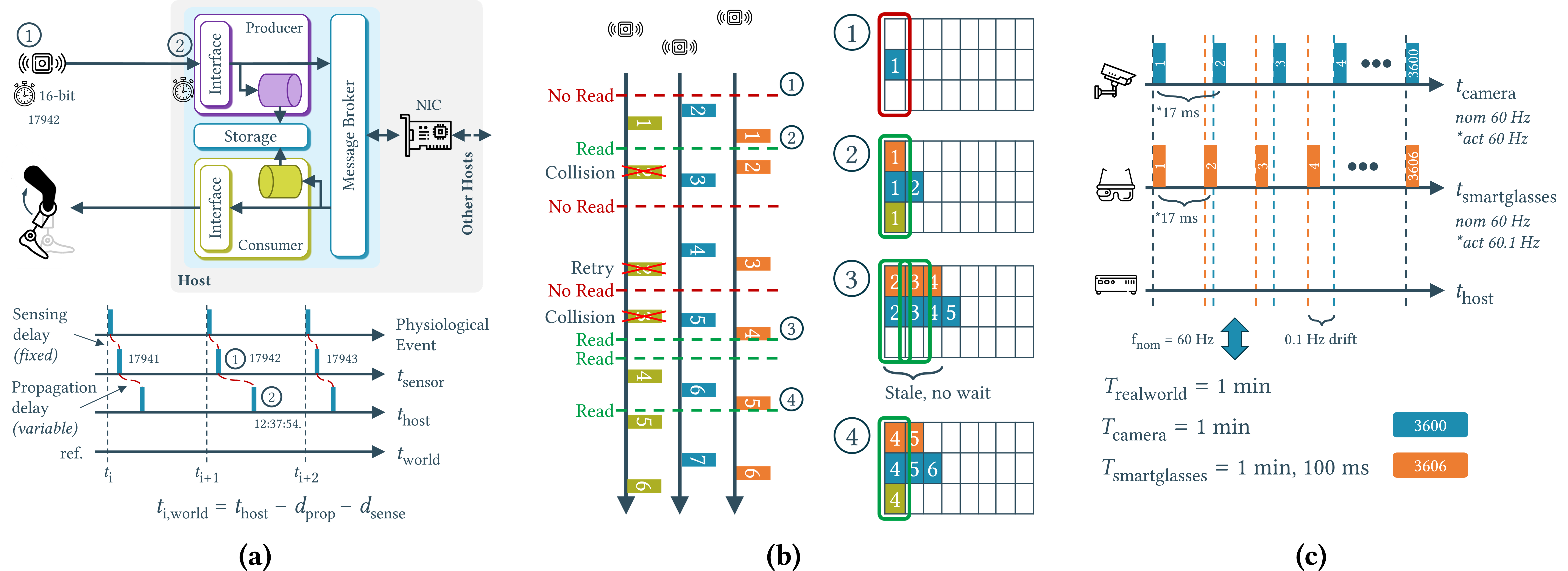}
  \caption{Flow path of a sensor sample through the system, stamped with the host's time on arrival (a).
  Inter-alignment of asynchronously received correlated measurements of distinct IMUs, experiencing congestion in wireless communication (b).
  Desynchronization between a 60 Hz sensor with an actual sample rate of 60.1 Hz, over a minute window, and a sensor with a perfect 60.0 Hz onboard clock: the former produces 6 samples more each minute (100 milliseconds worth), making direct one-to-one pairing of samples increasingly incorrect overtime, unless data is timed w.r.t. host's time (c).}
  \Description{
  In "a", .
  In "b", the timeline of arrival of samples from three separate IMU sensors, having occasional collisions and retransmission of samples.
  The alignment mechanism buffers arrived samples with respect to a user-defined parameter defining the period before a measurement is considered stale and is no longer to be awaited. The mechanism permits the upstream reader to read data when a measurement for each sensor is available, or if one of the sensors has not provided a valid sample for longer than the stale parameter.
  In "c", .
  }
  \label{fig:synchronization_details}
\end{figure*}

%% file: content/05-evaluation.tex
\subsection{Evaluation}
\label{sec:evaluation}
We followed the methodology of Sec.~\ref{sec:methodology} to contextualize the use case of distributed multimodal sensing for realtime intelligent prosthesis control, Fig.~\ref{fig:prosthesis_usecase}.
We used \texttt{HERMES} to reliably collect synchronized physiological data from 14 participants over hour-long experiments, and to characterize the selected hardware sensing setup in the natural environment to shape the design of a conceptual downstream AI model.
The closed-loop PyTorch integration used an untrained TCN classifier \cite{krug-pytorchtcn-github-2024} on raw sensed IMU data, to validate the core in-the-loop inference feature of the framework.

The selected use case emphasizes the complexity and the extent of heterogeneity in the real-world sensing and closed-loop AI-based control applications.
It is a concrete real-world sample that other researchers can relate to and use to extrapolate to own scenarios.
The setting displays how \texttt{HERMES} unifies the interaction in such applications and guides development of intelligent systems.

\subsubsection{Hardware Specification}
\label{sec:hardware_specification}
The use case targeted commercially available computing devices and sensors, Tab.~\ref{tab:resource_utilization}.
It consisted of an Axiomtek IPC962-525 edge server with an Intel Core i7-9700TE CPU, NVIDIA P4 GPU for video encoding, 32 GB of RAM, and 4-port Gigabit PoE NIC for 4x GigE $30 fps$ 1440p Basler a2A2590 cameras.
An Intel NUC minicomputer with an Intel Core i7-1360P CPU and 32 GB of RAM, used as a proof-of-concept wearable host.
Consumer-grade WiFi 7 TP-Link Deco BE65 access points.
An HP EliteBook 835 G11 laptop with an AMD Ryzen 7 PRO 8840U CPU and 16 GB of RAM.
A Raspberry Pi 5 embedded prosthesis controller, with a quad-core Arm Cortex A76 CPU, WiFi, and 8 GB of RAM.
The wearable, prosthesis, and laptop were connected wirelessly to the local $5 GHz$ network.
The edge server was wired to one of the access points, and acted as a local NTP clock reference that continuously synchronized other hosts within $2-3 ms$ accuracy.

The wearable connected over USB-C to the Pupil Labs Core eye tracking and egocentric view smartglasses, and an IPoverUSB tethered UDP connection to a smartphone receiving $100 Hz$ wireless Moticon OpenGo pressure insoles data.
It streamed both in realtime over UDP from the corresponding OEM app to the localhost: variable rate gaze, and $30 fps$ egocentric vision from the Pupil Capture app, and the $100 Hz$ insole data from the Moticon dekstop app, respectively.
A full-body 17-piece Xsens Awinda $60 Hz$ inertial MoCap setup and an 8-piece lower-body Cometa Systems PicoLite $2 kHz$ surface electromyography (sEMG) wearable setup were wirelessly connected to the corresponding OEM base station, both wired to the laptop.
The IMU MoCap data was streamed in realtime into \texttt{HERMES} from the Xsens MVN Analyze app using the Network Streamer protocol \cite{xsens-mvnetworkstreamer-xsens-2020} over UDP.
The sEMG data, streamed in bursts of 200 measurements every $100 ms$, was directly accessed via our custom Python binding to the low-level OEM SDK.
All network streams and SDKs provided realtime data, with varying internal latencies, from the heterogeneous distributed sensing setup to the device interface layer of \texttt{HERMES} at each corresponding host.

The embedded prosthesis controller operated embedded Linux, the rest of the hosts - Windows, to be compatible with the Windows-only OEM sensor interface software.
Hosts ran \texttt{HERMES} framework inside a Python 3.13 environment with FFmpeg 7.1.1, 30-second storage flush period on the wearable, prosthesis and laptop, and one-second period on the edge server.

\input{tables/resource_utilization}

\input{figures/prosthesis_usecase}

\input{figures/evaluation}

\subsubsection{Performance Results}
\label{sec:performance_results}
We conducted two key measurements of the overall system.
One, benchtop throughput and latency of \texttt{HERMES} on the selected hosts, Fig.~\ref{fig:evaluation}(a-b) - to determine system performance and suitability of the computing hardware for the use case.
Two, characterization of the sensed physiological signals in the natural environment, Fig.~\ref{fig:evaluation}(c-d), and complete system resource usage, Tab.~\ref{tab:resource_utilization} (column 2-3).
It facilitated the analysis of the nature of the real-world signals for shaping AI model design requirements, and offered processing headroom insights, respectively.

Performance evaluation showed consistent throughput scaling in sample rate up to $2kHz$ for fixed $1kB$ messages, Fig.~\ref{fig:evaluation}a and scaling with message size at a fixed $100Hz$ rate below $200kB$, Fig.~\ref{fig:evaluation}b.
The $2kHz$ cut-off is motivated by the inability of the test signal generation loop to produce large signals at such high rates.
Real-world characterization, Tab.~\ref{tab:resource_utilization} (column 2-3), sustained the four high-resolution videos at the edge server, saturating its resources: encoding consumed $14\%$ of CPU per video stream, the remaining $17\%$ - by our framework.
Majority of the CPU resources on the wearable were consumed by the smartglasses OEM interfacing app at $48\%$.
The prosthesis and laptop hosts with low-level networking and SDK integrations at the interface layer, and without video processing, retained a high headroom for additional processing.
Figure.~\ref{fig:evaluation}(c-d) offered an outlook on the real-world missingness of each modality that the downstream AI must be robust to.

The validation of the closed-loop AI inference functionality alongside sensing, in the prosthesis use case, showed a $4ms$ average and $2-5ms$ peak-to-peak processing latency with the non-optimized 32-bit floating point TCN model, contributing $12\%$ of CPU usage.

\subsubsection{Case Study Insight}
\label{sec:case_study_insight}
System profiling in Sec.~\ref{sec:performance_results} contextualized the constraints and the nature of the signals for the edge AI algorithm development.
It reproduced the observation of the natural missingness in the multimodal physiological data of Xu, et al. \cite{xu-lsm-arxiv-2025}, with a spread in reliability across modalities.
Regardless of the inference strategy of Sec.~\ref{sec:inference_strategies}, the designed model must be robust to prolong missingness in the less reliable modalities (insoles), and to the unstructured missingness of 
$P(X_i)=p_i$, $\mathcal{P} = \{p_i \in [0,1], 1 \leq i \leq N\}$,
where $N$ and $p_i$ the number of modalities and the rate of the corresponding modality's missingness.

Assuming a domain-informed movement intent forecasting horizon constraint of $290ms$ \cite{ma-rtlmr-jbhi-2025} for effective prosthesis actuation, the profiled $45ms$ total delay of the pilot setup and the $3ms$ message transmission delay to the prosthesis, the absolute maximum time budget for a task transition to be detected by the AI algorithm before a key event occurs \cite{wang-enhancedlmr-jbhi-2025} is $242ms$.
Following Sec.~\ref{sec:methodology}, three balancing flows become available in the following order, Fig.~\ref{fig:ai_design_context}. 
One, a balance between latency of the selected model, hence model size, and its forecasting time horizon.
Two, the \texttt{PUSH} or \texttt{PULL} inference strategy of Sec.~\ref{sec:inference_strategies}, impacting the time budget between consecutive inferences. 
Three, the size of overlap over the multi-variate time-series data, that retains AI model architecture and expands the inference time budget at the expense of sparser predictions.
Using these constraints, a potential AI model is parametrized, instantiated, and tested within the same sensing setup to estimate the new resource utilization, prior to model training and validation.

\input{figures/ai_design_context}

%% file: tables/resource_utilization.tex
\begin{table}
  \caption{Memory and CPU utilization of \texttt{HERMES} (excl. OEM apps and OS), and total device usage in parentheses.}
  \label{tab:resource_utilization}
  \begin{tabular}{l l c c c c}
    \toprule
    \multicolumn{2}{l}{\textbf{Device}}
    & \multicolumn{1}{c}{\textbf{Mem (GB)}}
    & \multicolumn{1}{c}{\textbf{CPU (\%)}}
    & \multicolumn{1}{c}{\textbf{Modal}}
    & \multicolumn{1}{c}{\textbf{Rate (Hz)}} \\
    \midrule
    \multicolumn{2}{l}{\textit{Edge server}}
    & 1.7 (7.6)
    & 73 (100)
    &
    \\
    & Cameras
    &
    &
    & 4
    & 30
    \\
    \multicolumn{2}{l}{\textit{Wearable}}
    & 0.6 (11.3)
    & 17 (68)
    &
    \\
    & Ego vision
    &
    &
    & 1
    & 30
    \\
    & Gaze
    &
    &
    & 4
    & 120
    \\
    & Insoles
    &
    &
    & 1
    & 100
    \\
    \multicolumn{2}{l}{\textit{Laptop}}
    & 0.3 (7.9)
    & 5 (16)
    &
    \\
    & IMU MoCap
    &
    &
    & 6
    & 60
    \\
    & sEMG
    &
    &
    & 1
    & 2000
    \\
    \multicolumn{2}{l}{\textit{Prosthesis}}
    & 0.6 (0.9)
    & 28 (41)
    &
    \\
    & Telemetry
    &
    &
    & 1
    & 10
    \\
    \bottomrule
  \end{tabular}
\end{table}

%% file: figures/prosthesis_usecase.tex
\begin{figure}
  \centering
  \includegraphics[width=0.8\columnwidth]{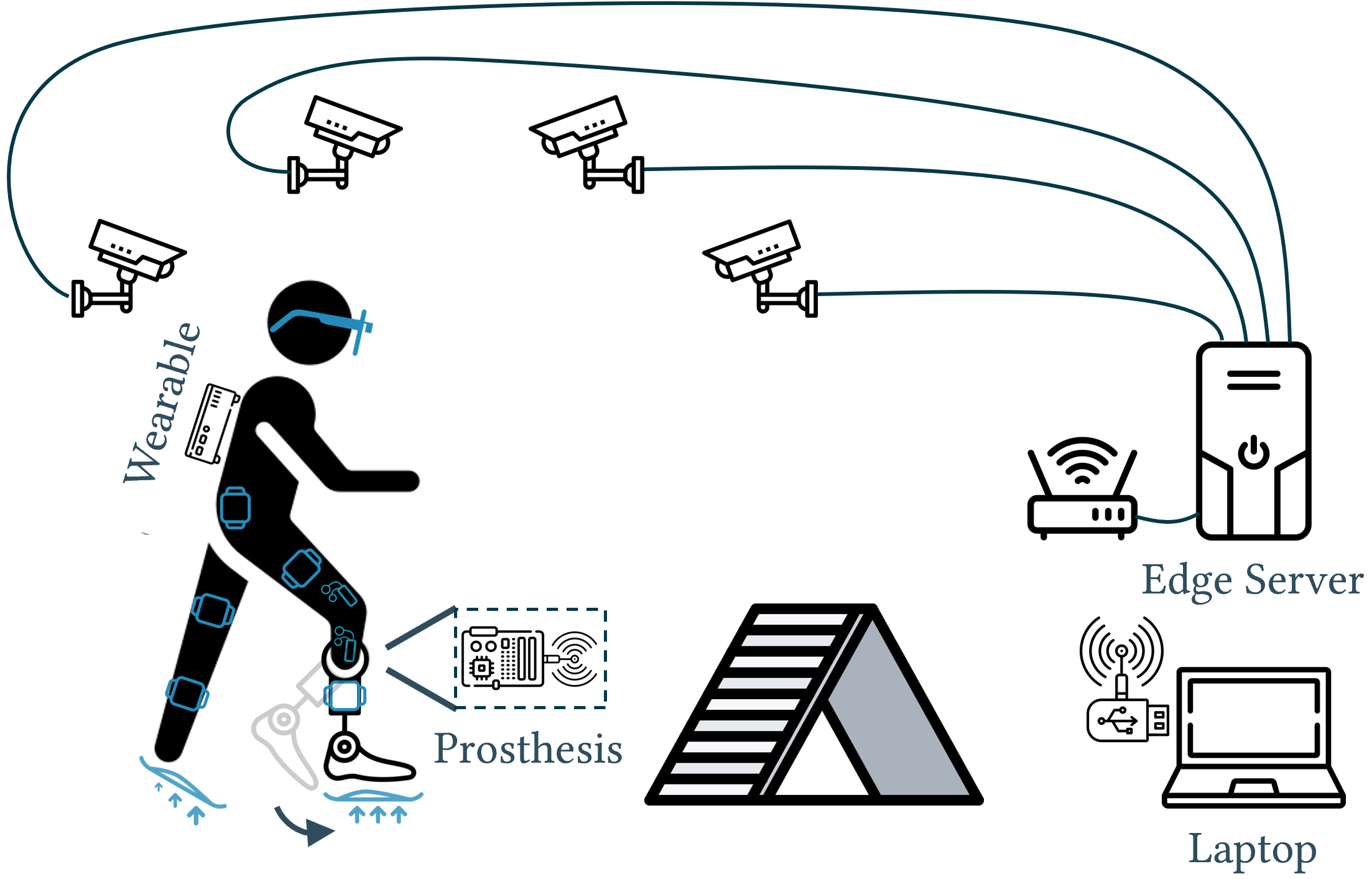}
  \caption{Usecase of closed-loop prosthesis control from wearable and externally captured data via AI-based intent prediction on the wearable single-board computer.}
  \Description{A pictogram visualizing the intelligent prosthesis application.
  The amputee approaches an obstacle with a powered connected prosthesis, wearing a set of physiological sensors and a wearable minicomputer on their back.
  Multiple physiological sensors are visible on the person's body - inertial sensors, electromyography, and foot pressure sensors on the lower extremities, and smartglasses on the person's face for gaze tracking and egocentric vision.
  Externally, the person is recorded with four synchronized high-resolution cameras, wired to the edge server, which is in turn connected to the rest of the host devices through a wireless access point for communication and realtime exchange of sensor information.
  The prosthesis is shown to trigger knee extension as the person approaches the obstacle.
  The sensing setup is monitored remotely via a researcher's laptop, connected to the same network.}
  \label{fig:prosthesis_usecase}
\end{figure}

%% file: figures/evaluation.tex
\begin{figure*}[t]
  \centering
  \includegraphics[width=\linewidth]{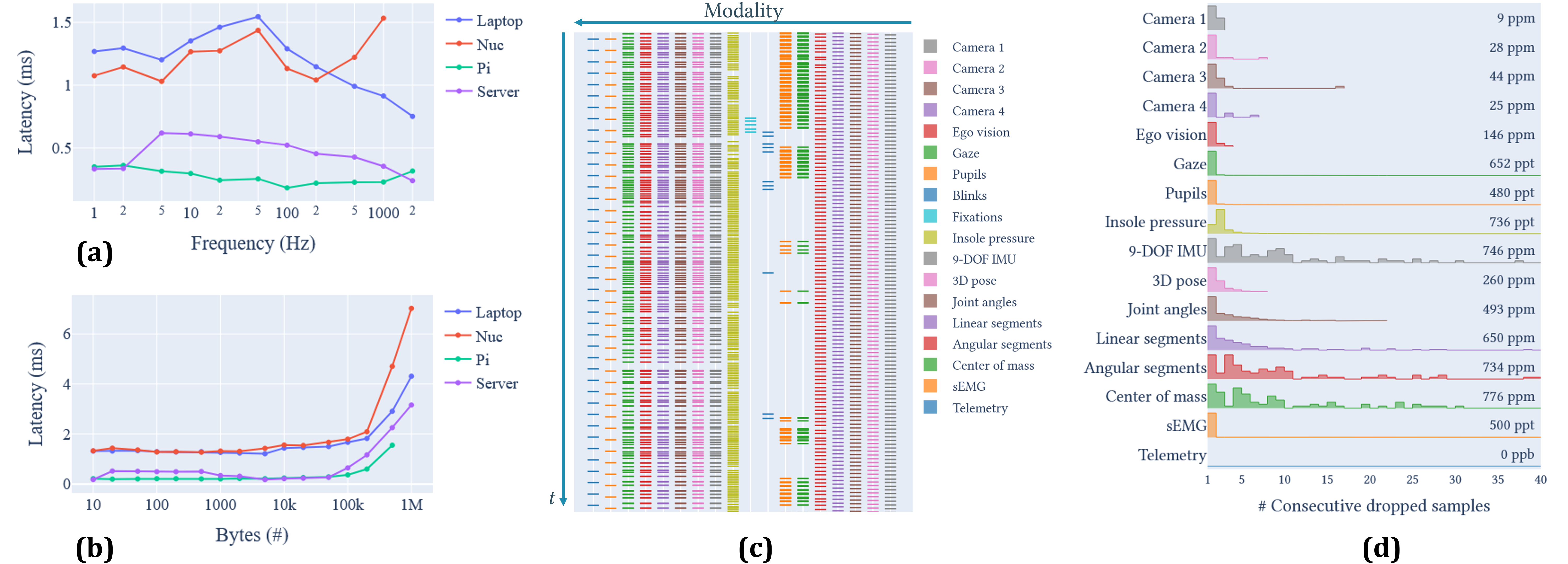}
  \caption{Overall system performance overview on the intelligent prosthesis use case of Fig.~\ref{fig:prosthesis_usecase}.
  Latency w.r.t. the message size (a) and w.r.t. the sample rate (b).
  Overview of constant rate lossy, and event-driven modalities (c).
  Distribution of the missingness durations across modalities in part per notation, recorded for 14 subjects over $0.7 \pm 0.1$ hour windows each (d).}
  \Description{
  In "a", a line plot of some of the hosts performance with variable sample rate, for a message of fixed size that decreases gradually due to internal optimization of ZeroMQ. 
  In "b", a line plot of some of the hosts performance with variable message size at a fixed sample rate, showing consistent latency scaling with increasing throughput until a saturation point.
  In "c", an overview of the asynchronous nature across modalities, displayed as a grid plot. On the horizontal axis - modalities of the prosthesis use case, on the vertical - arrival time of each corresponding sample of that modality. The plot emphasizes the asynchrony between modalities, even though a sensor is monotonically sampling a signal. Pupil, gaze, blinks, and fixation detection are event-based and have non-deterministic gaps between consecutive samples. All 4 are derived from the egocentric and eye facing cameras of the smartglasses, and their reliability is based on it. 
  In "d", a ridgeline plot of missingness across modalities in the real-world validated setup of the intelligent prosthesis use case. Listing on the vertical axis modalities, and on the horizontal axis - histogram of missingness events, with respect to their duration. Each trace is annotated with an overall parts-per-million, parts-per-thousand, or parts-per-hundred count of total missed samples. Each trace has a different length tail, which indicates the spread of observed missingness events: the wider the tail, the less reliable the modality. The plot shows that fixed cameras and egocentric vision are generally very reliable, while other wearable modalities exhibit occasional prolong disconnection periods. Insoles have long disconnection periods, IMU MoCap modalities have diverse missingness durations, and telemetry has no missingness because data is generated onboard the prosthesis itself.
  }
  \label{fig:evaluation}
\end{figure*}

%% file: figures/ai_design_context.tex
\begin{figure}[ht]
  \centering
  \includegraphics[width=0.8\columnwidth]{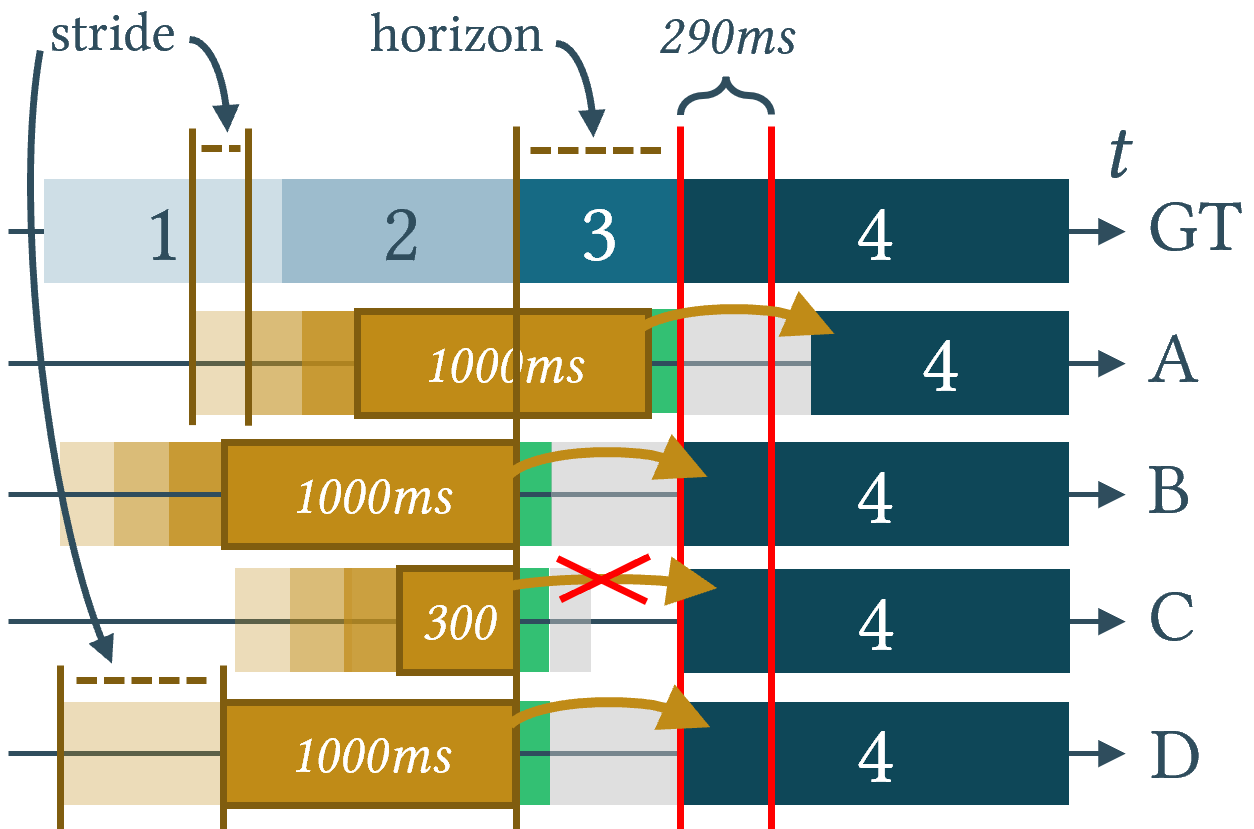}
  \caption{Context guided AI model design flow for time-limited detection of actions 1-4 (GT), over overlapping windows of data (brown) with total internal sensing delay (green) and processing latency (gray). Initial detection model with no forecasting (A). Model with sufficient forecasting but latency larger than sample period (B). Realtime model with insufficient forecasting (C). Fitting model with larger stride and sparse predictions (D).}
  \Description{xxxxxxxxxxxxxxxxxxxxxx}
  \label{fig:ai_design_context}
\end{figure}

%% file: content/06-discussion.tex
The low CPU and memory footprint of the framework, Tab.~\ref{tab:resource_utilization} (column 2-3), stable steady-state memory profile, and scaling performance, Fig.~\ref{fig:evaluation}(a-b), confirm that our framework was implemented in a highly efficient parallelized way to handle large volume of physiological wearable and external data.
Decreasing latency with higher sample rates in Fig.~\ref{fig:evaluation}a is explained by the internal optimizations of ZeroMQ \cite{hintjens-zeromq-2013} that batches transactions for better and more responsive performance.
The footprint of our framework on the embedded prosthesis host contributed $3\%$ of the total CPU and $100MB$ of memory usage, with the remainder consumed by the control algorithms, illustrating the framework's resource efficiency.

Measure of missingness in the target environment, Fig.~\ref{fig:evaluation}(c-d), provides insight on the distribution of durations of naturally missing data events: left - random single lost samples, right - prolong disconnection events.
The narrower the distribution, the higher the number of the corresponding events.
The wider the tail, the higher is the spread of the missingness durations and hence the less reliable the modality.
The the parts per notation on each modality, Fig.~\ref{fig:evaluation}d, shows the ratio of overall missing samples to complement the insight on duration of the events.  
The figure shows a high data reliability across modalities, each marked with the percentage of time a sample was missing. 
Except for pressure insoles that had prolong rare disconnection periods of $\sim3min$, and of $\sim2-3s$ for sEMG, other wearable and external modalities had random short missingness events not exceeding $40$ consecutive samples, and $0$ missingness for telemetry data generated onboard the prosthesis.
Event-driven modalities, like the detected fixations and blinks, do not have an inherent missingness, but are asynchronous in nature.
Pupil detection and gaze information are derivatives of the $120Hz$ eye facing cameras, detection algorithm, and human eye movement.
Both are displayed as constant rate modalities on Fig.~\ref{fig:evaluation}d, but are in fact asynchronous event-based modalities as illustrated by Fig.~\ref{fig:evaluation}c.

OEM sensor interface apps consume a large portion of system resources, followed by video encoding, exceeding the footprint of the framework itself.
Overall system performance is tightly coupled to the underlying hardware, making it challenging to determine hardware suitability upfront.
Characterization of a complete setup in the target environment is more reliable and displays the computational intensity of the in-the-loop AI processing component that can fit in the application, pointing at a hardware upgrade or a processing algorithm downgrade requirement.
Replacing the outdated edge server CPU with a modern Intel Core i7-14700T alternative after observations from the sample use case, facilitates realtime video encoding of the high-resolution videos with $0$ missingness on the CPU alone, and reduces CPU usage to $46\%$.

Our "no captured data lost" policy guarantees that no samples captured by the host are dropped due to slow consumption, operating system scheduling, or frameworks internal processing.
This may result in memory overflow if the selected hardware is unsuitable, if the user's custom node logic improperly manages the data, or if downstream AI algorithm has higher latency than the data sample rate.
This can be addressed by fixing the maximum size of the \texttt{Stream} datastructure: it will prevent memory overflow, but at the cost of lost samples if the computing hardware is slow or the AI model is too computationally intensive.

Applying our proposed methodology and framework on the example intent prediction use case, uncovered real-world constraints for the downstream edge AI model selection. 
The real-world evaluation of the holistic system highlighted what sensing delay, modality missingness, an application must naturally tolerate and what headroom remains for processing.
The insights provided actionable guidelines toward the design and verification flow of an effective realtime closed-loop system.
Validation of the closed-loop processing capability of the present framework with the streamable TCN classifier is orthogonal to the methodology and was used solely to illustrate the framework's claimed core functionality.

The combination of the low-latency transport, optimized datastructures, and efficient processing, streamlined in-the-loop AI inference, allows \texttt{HERMES} to fill the unaddressed gaps in the domain by other SoA frameworks.
It satisfies the needs of researchers in the intelligent physiological applications for high-quality data collection and closed-loop realtime edge AI interventions. 



%% file: content/07-limitations.tex
The system is not medically certified and doing so falls outside of the scope.
It is not feasible to certify the framework for the exhaustive list of current and future commercial and custom sensors, actuators, and AI models. Some edge AI use cases target standalone extreme embedded devices - microcontrollers, that do not directly support high-level programming languages and networking libraries.
Integrating support for microcontrollers in future works will bring the present methodology into prolong daily-use low-power wearables.

The current implementation assumes a semi-controlled experiment, where all distributed hosts are launched individually, and terminated centrally at a coordinator host.
Cameras may overheat, OEM sensor interfacing software may crash, local network may be down.
The always-on free-living systems for at-home monitoring or for daily outdoor use require a fault-tolerant design, where any host can drop out and rejoin dynamically without a centralized coordinator.
Future extension of the framework with dynamic discoverability of free-running hosts will enable resilient always-on systems for long-term longitudinal studies in the wild.

Use case specific tightness of required inter-modality synchronization remains an open question.
Empirical evidence of the impact of different temporal misalignment in replayed real-world acquired data will present insights of the extent of inter-modality synchronization to which the requirements can be relaxed while preserving acceptable edge AI performance.

%% file: content/08-conclusion.tex
We proposed a unified methodology for tackling continuous realtime multimodal sensing and AI inference edge systems for the real-world, and an open-source framework that embodies it.
We demonstrated on a pilot use case that a holistic system-level approach complemented by \texttt{HERMES} contextualizes the requirements in this multifaceted challenge, shapes practical constraints, and guides the user around them toward a suitable closed-loop AI inference model.
We validated our framework on a relevant pilot use case to show its high-throughput low-latency capacity.
We opened it to the research community to lower the implementation barriers and to accelerate the translation of research efforts of the field into real-world edge AI healthcare applications.
We demonstrated how following the proposed methodology creates an abstraction from the technical challenges and enables the development of novel closed-loop intelligent applications.
We acknowledged the current limitations of the proposed framework and a practical path forward to addressing it in future works.
The work is far from the universal standardization of continuous realtime distributed multimodal edge AI, but is a practical publicly available attempt to converge to one.
All sourcecode is made openly available for reuse and reproduction.